\documentclass[11pt,a4paper]{article}
\usepackage{jcappub}

\title{Interacting agegraphic dark energy models in phase space}

\author{O.\,A.\,Lemets}
\author{D.\,A.\,Yerokhin}
\author{L.\,G.\,Zazunov}
\affiliation{Akhiezer Institute for Theoretical Physics,
National Science Center "Kharkov Institute of Physics and
Technology", Akademicheskaya Str. 1, 61108 Kharkov, Ukraine}
\emailAdd{oleg.lemets@gmail.com}
\emailAdd{denyerokhin@gmail.com}

\abstract{
Agegraphic dark energy, has been recently proposed, based on the so-called Karolyhazy uncertainty relation, which arises from quantum mechanics together with general relativity. In the first part of the article we study the original agegraphic dark energy model by including the interaction between agegraphic dark energy and pressureless (dark) matter. The phase space analysis was made and the critical points were found, one of which is the attractor corresponding to an accelerated expanding Universe.

Recent observations of near supernova show that the acceleration of Universe decreases. This phenomenon is called the transient acceleration. In the second part of Article we consider the 3-component Universe composed of a scalar field, interacting with the dark matter on the agegraphic dark energy background. We show that the transient acceleration appears in frame of such a model. The obtained results agree with the observations.}
\arxivnumber{1010.0567}
\begin{document}

\maketitle
\flushbottom
\toccontinuoustrue
\section{Introduction}
Nowadays it is strongly believed that the Universe is experiencing an accelerated expansion,
and this opinion is supported by many cosmological observations  such as SNe Ia \cite{Supernova}, WMAP \cite{WMAP}, SDSS \cite{LSS} and X-ray \cite{X-ray}. These observations suggest that the Universe is dominated by dark energy with negative pressure, which provides the dynamical mechanism for the accelerating expansion for the Universe. 

The simplest candidate of dark energy is a tiny positive cosmological constant. However, as is well known, it suffers from the so-called cosmological constant  \cite{dark_energy1} and coincidence problem \cite{ZWS}. The cosmological constant problem is essentially a quantum gravity problem, since the cosmological constant is commonly considered as the vacuum expectation value of some quantum fields. Since a completely successful quantum theory of gravity is still unavailable, it is more realistic directly to combine quantum mechanics with general relativity. 

Although the nature and origin of dark energy could
perhaps understood by a fundamental underlying theory unknown up to
now, physicists can still propose some paradigms to describe it. In
this direction we can consider theories of modified gravity
\cite{Bergmann,Odintsov}, or field models of dark energy. The field models
that have been discussed widely in the literature consider a
cosmological constant \cite{cosmo}, a canonical scalar field
(quintessence) \cite{quint}, a phantom field, that is a scalar field
with a negative sign of the kinetic term \cite{phant,phantBigRip}.

However, we still can make some attempts to probe the nature of dark energy according to some
principles of quantum gravity although a complete theory of quantum gravity is not available today. The holographic dark energy model is just an appropriate and interesting example, which is constructed in the light of the holographic principle of quantum gravity theory \cite{Susskind,Hooft,Witten}. Before a completely  successful quantum theory of gravity is available, it is more  realistic to combine quantum mechanics with general relativity  directly.

Following the line of quantum fluctuations of spacetime, in Refs.~\cite{Sasakura,Maziashvili_1,Maziashvili_2,0707.4049},
 we use the so-called K\'{a}rolyh\'{a}zy relation~\cite{Karolyhazy}
 \begin{equation}
\label{eq0}
 \delta t = \beta t^{2/3}_p t^{1/3},
 \end{equation}
 where $\beta$ is a numerical factor of order one,  $t_p$ is the  reduced Planck time, and throughout this paper, we use the units $c=\hbar  =k_b=1$, so that one has $l_p=t_p=1/M_p$ with $l_p$ and $M_p$ being the  reduced Planck length and mass, respectively.

The K\'arolyh\'azy relation (\ref{eq0}) together with the
time-energy uncertainty relation enables one to estimate a quantum
energy density of the metric fluctuations of Minkowski
space-time~\cite{Maziashvili_1,Maziashvili_2,0707.4049}. With the relation (\ref{eq0}), a length scale $t$ can be known with a maximal precision $\delta t$
determining a minimal detectable cell $\delta^3$ over a spatial
region $t^3$. Thus one is able to look at the microstructure of
space-time over a region $t^3$ by viewing the region as the one
consisting of cells $\delta t^3 \sim t_p^2t$. Therefore such a cell
$\delta t^3$ is the minimal detectable unit of space-time over a
given length scale $t$ and if the age of the space-time is $t$, its
existence due to the time-energy uncertainty relation cannot be
justified with energy smaller than $\sim t^{-1}$ . Hence, as a
result of the relation (\ref{eq0}), one can conclude that if the age
of the Minkowski space-time is $t$ over a spatial region with linear
size $t$ (determining the maximal observable patch) there exists a
minimal cell $\delta t^3$, the energy of the cell cannot be smaller
than~\cite{Sasakura,Maziashvili_2,0707.4049}
 \begin{equation}
  \label{eq02}
  E_{\delta t^3} \sim t^{-1},
  \end{equation}
  due to the time-energy uncertainty relation, it was argued  that the energy density of metric fluctuations of  Minkowski spacetime is given by
 \begin{equation}
\label{eq1}
 \rho_q\sim\frac{E_{\delta t^3}}{\delta t^3}\sim
 \frac{1}{t_p^2 t^2}\sim\frac{M_p^2}{t^2}.
\end{equation}

In~\cite{Maziashvili_2} (see also~\cite{0707.4049}), it is noticed that the
 K\'{a}rolyh\'{a}zy relation naturally obeys the  holographic black hole entropy bound \cite{Bekenstein}.  It is worth noting that  the form of energy density eq.~(\ref{eq1}) is similar to the one  of holographic dark energy~\cite{0707.2129,0701405}, i.e.,  $\rho_\Lambda\sim l_p^{-2}l^{-2}$. The similarity between  $\rho_q$ and $\rho_\Lambda$ might reveal some universal features  of quantum gravity, although they arise from different sources.
As the most natural  choice, the time scale $t$ in eq.~(\ref{eq1}) is chosen to
 be the age of our Universe. Therefore, we call it ''agegraphic'' dark  energy \cite{0707.4049}. 

There are many variations of models with agegraphic dark energy (ADE) to solve most problems of SCM. Particularly,  various articles consider the possibility of interaction of this kind of dark energy with dark matter \cite{0707.4052,shey1}.

Some authors have recently suggested that the cosmic acceleration have already peaked and that we are currently observing its slowing down \cite{Barrow,Starobinsky,Lima}.
Under a kinematic analysis of the most recent SNe Ia compilations, in the paper \cite{Lima}  shows  the existence of a considerable probability in the relevant parameter space that the Universe is already in a decelerating expansion regime.

We does not know  works, which investigated the possibility to explain this phenomenon in the framework of ADE model.
Thus, it is interesting to consider the model with the interacting dark energy, for transient acceleration. Typically, in models with transient acceleration the role of dark energy played by scalar field, non-monotonic dynamics of which provide the necessary dependence of deceleration parameter. In our work we made an attempt to combine the scalar field model with agegraphic dark energy.

The paper is organized as follows: in the next section we consider the dynamic features of the standard agegraphic dark energy interacting with the substance with an arbitrary equation of state $w.$ In section \ref{ISFA_TA} we investigate a new quintessence scenario driven by a rolling homogeneous scalar field in the exponential potential interacting with dark matter on the agegraphic background. It will be shown that this scenario predict transient accelerating phase. Consistency of these models with the observational data will be shown in Secs. \ref{OBS}.
\section{Review on the interacting  ADE}
In this Section we  consider the dynamics of the three component  Universe, which consists of ADE, dark matter and radiation. Since the nature of dark energy is not fully known, we can assume that the dark energy and dark matter interact.
 The energy density of ADE is given by \cite{0707.4049}
\begin{equation}
  \label{ADE}
  \rho_q=\frac{3n^2M_p^2}{T^2}.
\end{equation}
Here $M_p=(8\pi G)^{-1/2}$ and $T$ is chosen to be the age of our
Universe
\begin{equation}
  \label{T}
  T=\int^t_0{dt'}=\int^a_0{\frac{da}{Ha}},
\end{equation}
where $a$ is the scale factor of our Universe, $H\equiv\dot{a}/a$ is
the Hubble parameter and the dot denotes the derivative with respect
to cosmic time. 

Considering the flat Friedmann-Robertson-Walker Universe with the
agegraphic dark energy and substance with an arbitrary equation of state $w$ and energy density $\rho_w,$ the corresponding Friedmann equation is
\begin{equation}
  \label{fridm}
  H^2=\frac{1}{3M_p^2}(\rho_q+\rho_w+\rho_\gamma).
\end{equation}

The conservation laws of the ADE
and matter are respectively \cite{0707.4052}

\begin{eqnarray}
     \nonumber
  \dot{\rho}_w+3H(1+w)\rho_w&=&Q,\\
\label{cons_eq}
  \dot{\rho}_q+3H(1+w_q)\rho_q&=&-Q,\\
\dot{\rho}_\gamma+4H\rho_\gamma&=&0,
\end{eqnarray}

where $Q$ denotes the phenomenological interaction term. Note that the interaction is introduced in the form that satisfies covariance of the total energy momentum tensor. In the literature most often one can find the fallowing forms of interaction \cite{0707.4052,Chimento,InteractionQ1,InteractionQ2,InteractionQ3,Wei,ZKGuo}
\begin{equation}
  Q=3\alpha H\rho_q,\quad 3\beta H\rho_w,\quad 3\gamma H(\rho_w+\rho_q),
\end{equation}
where $\alpha$, $\beta$ and $\gamma$ are positive constant
parameters.  We consider the most general of above cited types of interaction  
\begin{equation}
  \label{Q}
  Q= 3H(\alpha\rho_q+\beta \rho_w).
\end{equation}
\subsection{A phase-space view on the interacting ADE}
In this section, we investigate the global structure of the dynamical system via phase plane
analysis and compute the cosmological evolution by numerical analysis. We introduce the following variables:
\begin{eqnarray} 
  x=\frac{1}{M_pH}\sqrt{\frac{\rho_{_{q}}}{3}},\quad y=\frac{1}{M_pH}\sqrt{\frac{\rho_{_{w}}}{3}},\quad z=\frac{1}{M_pH}\sqrt{\frac{\rho_{\gamma}}{3}}.\label{var1}
\end{eqnarray}
Taking the derivative of both sides of the Friedman equation (\ref{fridm}) with respect to the logarithm of the scale factor, $N = \ln a,$ and using eqs. (\ref{cons_eq}), it is easy to find that 
\begin{equation}
\label{dH_xyz}
    \frac{dH}{dN} =- \frac{3H}{2}\left((1+w_q)x^2+(1+w)y^2+\frac{4}{3}z^2\right).
\end{equation}
Using (\ref{var1}) and (\ref{dH_xyz}) the governing equations of motion (\ref{cons_eq}), (\ref{ADE})  could be reexpressed as the following system of equations:
\begin{eqnarray} 
x'&=& \frac{3x}{2}f(x,y,z)-\frac{x^2}{n},\nonumber\hfill\\
 \label{sys_xyz_1}
y'&=& \frac{3y}{2}f(x,y,z) -\frac{3}{2}(1+w)y,\hfill\\
z'&=& \frac{3z}{2}f(x,y,z) -2z,\nonumber\hfill
\end{eqnarray}

where 
$$
f(x,y,z)=\frac{2}{3n}x^3-\alpha x^2 +(1+w-\beta)y^2+\frac{4}{3}z^2
$$
and the prime denotes a derivative with respect to the
logarithm of the scale factor, $N = \ln a.$
The  first Friedmann equation can be written as
\begin{equation} \label{xyz}
    x^2+y^2+z^2 = 1.
\end{equation}
 Since it is limited by the condition $\rho >0$ to the circle 
$x^{2}+y^{2}+z^{2}\leq 1$, we limit the analysis only to the quarter of
unitary sphere of positive $x,y,z$. 

The state parameter $w_q$ for the agegraphic dark energy  could be expressed in terms of the new variables as
\begin{equation}
   \label{w_q_xy}
w_q=-1-\alpha+\frac{2}{3n}x-\beta\frac{y^2}{x^2}.
\end{equation}
The total equation of state parameter is given by 
\begin{equation}
w_{tot} =\frac{p_{tot}}{\rho_{tot}}=\frac{p_{q}+p_{w}+p_{\gamma}}{\rho_{q}+\rho_{w}+\rho_{\gamma}}=
\frac{w_qx^2+wy^2+\frac{1}{3}z^2}{x^2+y^2+z^2}.
\end{equation}
For completeness, we give the deceleration parameter 
\begin{equation}
    \label{q_}
q=-\frac{\ddot{a}}{aH^2}=-1-\frac{\dot{H}}{H^2}.
\end{equation}
Combining it with the Hubble parameter and  noting that $\dot{H} = H'H $ we obtain the deceleration parameter in terms of the (\ref{var1}) variables
\begin{equation}
q=-1+\frac 32\left[\frac{2}{3n}x^3+\alpha x^2 +(1+w-\beta)y^2+\frac{4}{3}z^2 \right] 
\end{equation}
It is worth noting that in the absence of interaction between  agegraphic dark
energy and dark matter, $\alpha=\beta=0,$ from eq. (\ref{w_q_xy}) we see that
$w_q$ is always greater than $-1.$  In addition the condition $n>1$ is necessary to derive the present accelerated expansion. However, the situation change as soon as the interaction term is taken into account.

In interacting agegraphic models of dark energy, the properties of agegraphic dark energy are determined by the parameters $n,\alpha$ and $\beta$ together.

\subsection{Phase plane analysis} 
\begin{figure}
\centering
\includegraphics[width=0.5\textwidth]{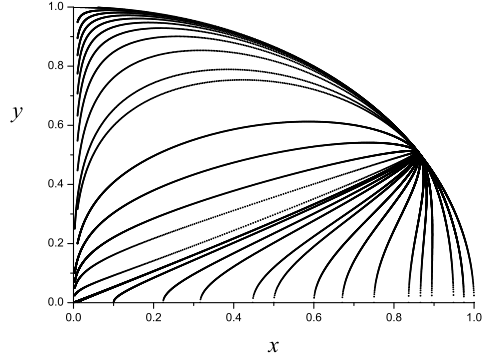}
\caption{The phase plane for $n = 3,\;\alpha = 0.25$ and $\beta= 0.1.$ The late-time attractor is the scaling solution with $x_* \approx 0.86,\; y_* \approx 0.26\; (\Omega_{q* } \approx 0.74,\;\Omega_{m* } \approx 0.26).$}
\label{fig:1}
\end{figure}

As said above, the motivation for interacting agegraphic models is to solve or, at least, ameliorate the coincidence problem. 
The goal will be achieved if the system (\ref{sys_xyz_1}) presents scaling solutions. As already discussed, scaling solutions are characterized by a constant dark matter to dark energy ratio $r=y^2/x^2.$ It is easy to find that
\begin{equation}
    r' = 3\alpha -3(1+w-\beta-\frac{2}{3n}x)r
\end{equation}
The scaling solutions mean $r'=0,$ as a result $r_{_*}=\alpha(1+w-\beta-\frac{2}{3n}x_{_*})^{-1}.$
On the other hand in this mode $x'= y'=0, \; z\approx 0$ and  we obtain from the first
equation in (\ref{sys_xyz_1}) that
\begin{equation}\label{x_att}
  x_{_*}=\frac{1}{2}\left(-b+\sqrt{b^2+4}\right), 
\end{equation}
where $b=\frac{3n}{2}\left((1+w-\beta)r_{_*}-\alpha\right).$ For the case shown in figure  \ref{fig:1}, we find $r_{_*} \approx 0.353$ and accordingly $x_{_*} \approx 0.86.$

Even more important are those scaling solutions that are also an attractor. In
this way, the coincidence problem gets substantially alleviated because, regardless of the initial conditions, the system evolves toward a final state where the ratio of dark matter to dark energy stays constant (see figure \ref{fig:1}).
The location of attractor, as shown above (\ref{x_att}), depends on the values $\alpha$ and $\beta.$ The simplest finite critical points and their properties for this model are summarized   in Table \ref{tab:1}.

\begin{table}
\caption{Location of the critical points of the autonomous system of eqs. (\ref{sys_xyz}), their stability and dynamical behavior of the Universe at those points.}
\label{tab:1}       
\begin{tabular}{cccc}
\hline\noalign{\smallskip}
$(x_c,y_c,z_c)$ & Stability & $q$ & $w_q$ \\
 coordinates &  character  & &  \\
\noalign{\smallskip}\hline\noalign{\smallskip}
$\;(0,0,1)$  &  unstable & 1 &$ -1-\alpha $\\
$(0,1,0)$ &unstable &\quad$-1+\frac{3}{2}(1+w-\beta)$ &$\nexists$\\
 $(x_*,y_*,0)$ &attractor&\quad$q_*<0$&$w_{q*}$\\
\noalign{\smallskip}\hline
\end{tabular}
\end{table}

The first critical point, $(0, 0, 1),$ is unstable and  corresponds to the radiation dominated era.   The other point, $(0, 1, 0),$ is physically unrealistic. It corresponds to a Universe filled with  matter with parameter of state $w$ (it contains neither radiation, nor dark energy) and it is also unstable.
Finally, the third critical point, $(x_*,y_*,0),$ is stable and in a natural way, the dynamical evolution of the Universe  asymptotically approaches this attractor, a never-ending phase of accelerated expansion, in which the energy ratio $r={y^2}/{x^2}$ remains constant (for $\alpha \neq 0$).
To conclude, the succession of expansion eras in the model under consideration can be summarized as follows: if the initial conditions satisfy $z_0>x_0 \neq 0,$ then  after the initial radiation dominated era, the Universe might have gone through a very short period of matter domination with  state parameter $w$ (if any) corresponding to an unstable critical point, to finally asymptotically approach a regime with a constant dark matter to dark energy ratio, at late times, in the accelerated expansion era.
Note that if initially the Universe does not contain dark energy then in the further evolution of dark energy does not exist. This fact is the fundamental difference from the models, in which the role of dark energy is played by scalar field \cite{0801.1565}.

\begin{figure}
\centering
\includegraphics[width=0.32\textwidth]{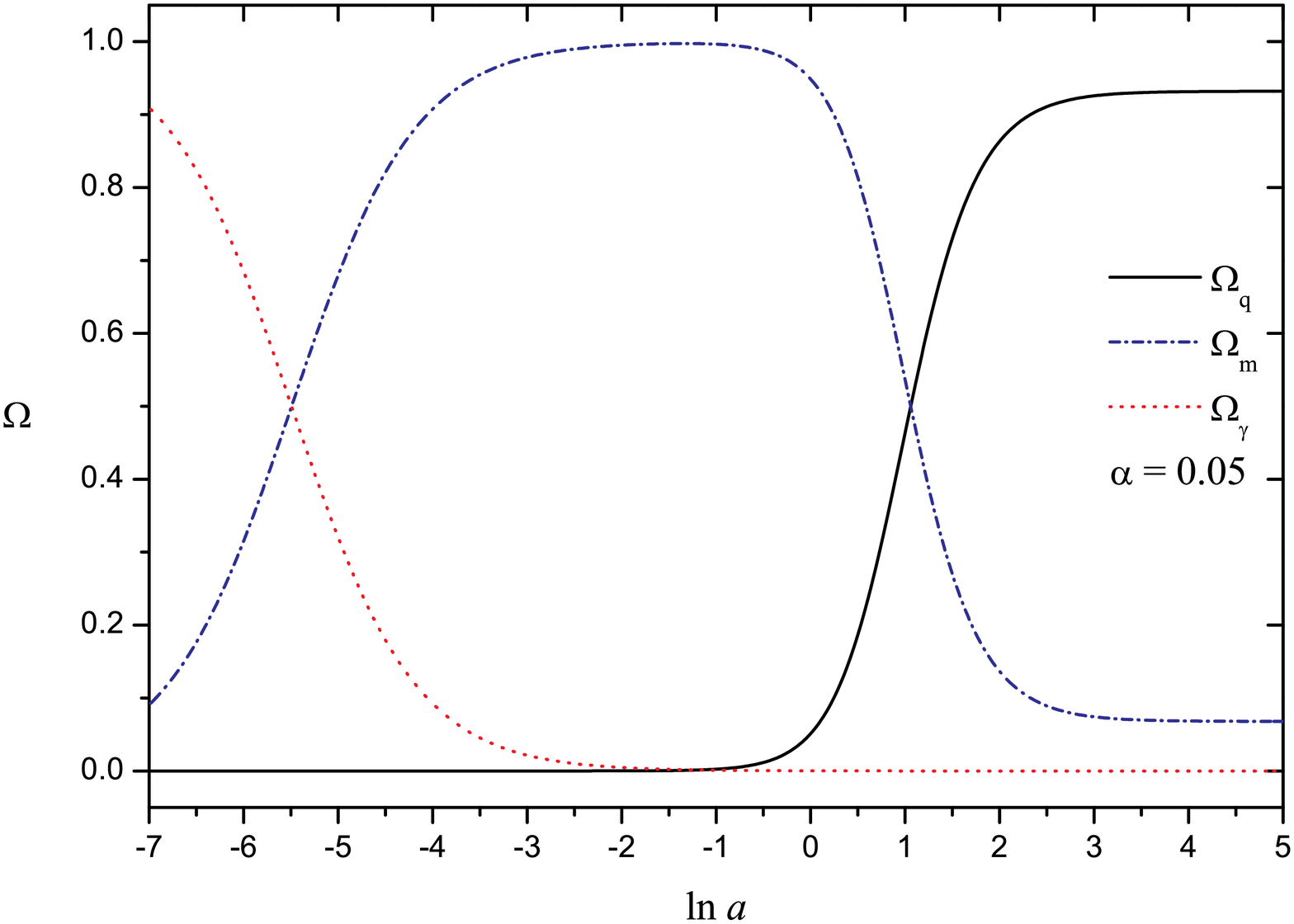}
\includegraphics[width=0.32\textwidth]{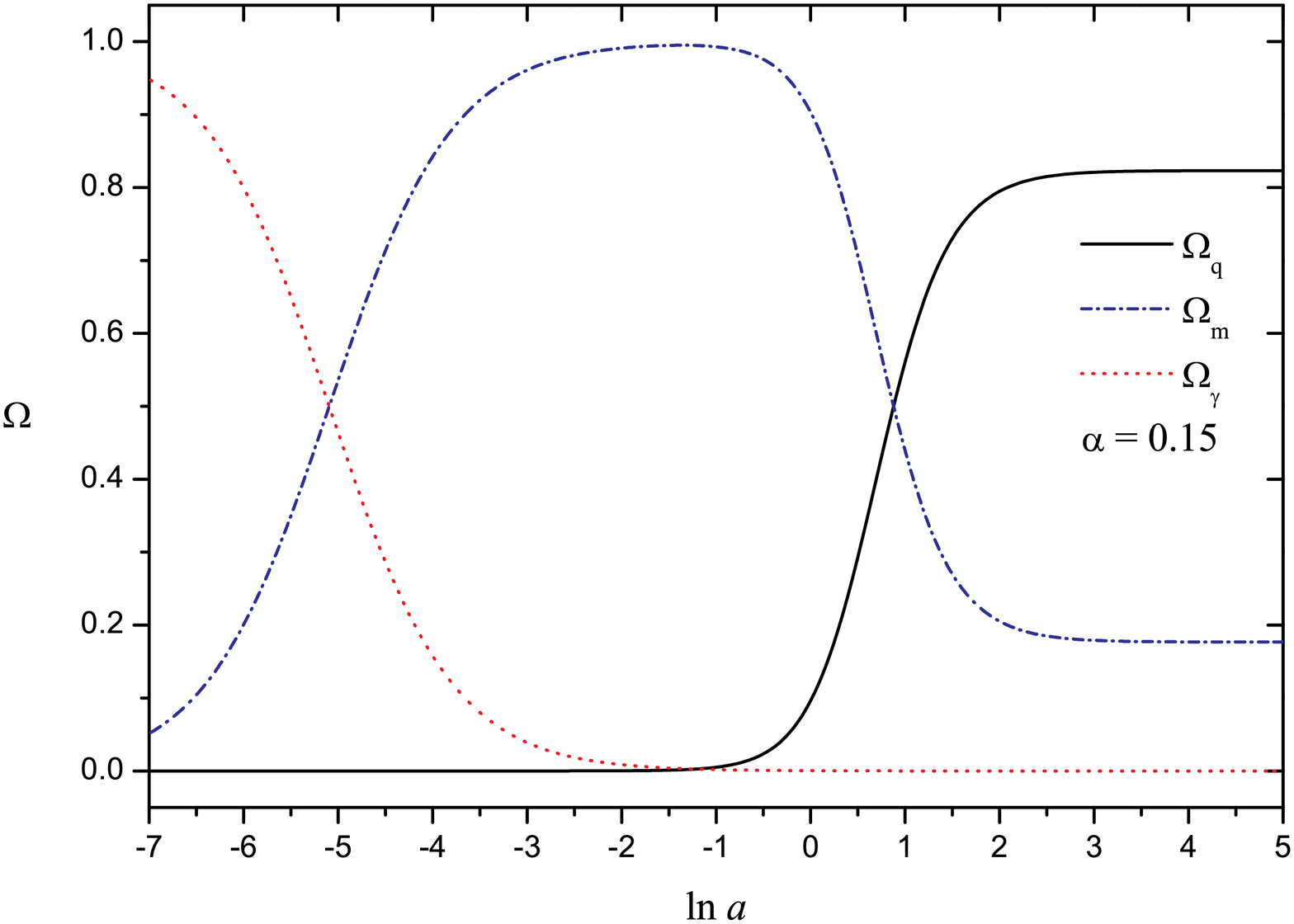}
\includegraphics[width=0.32\textwidth]{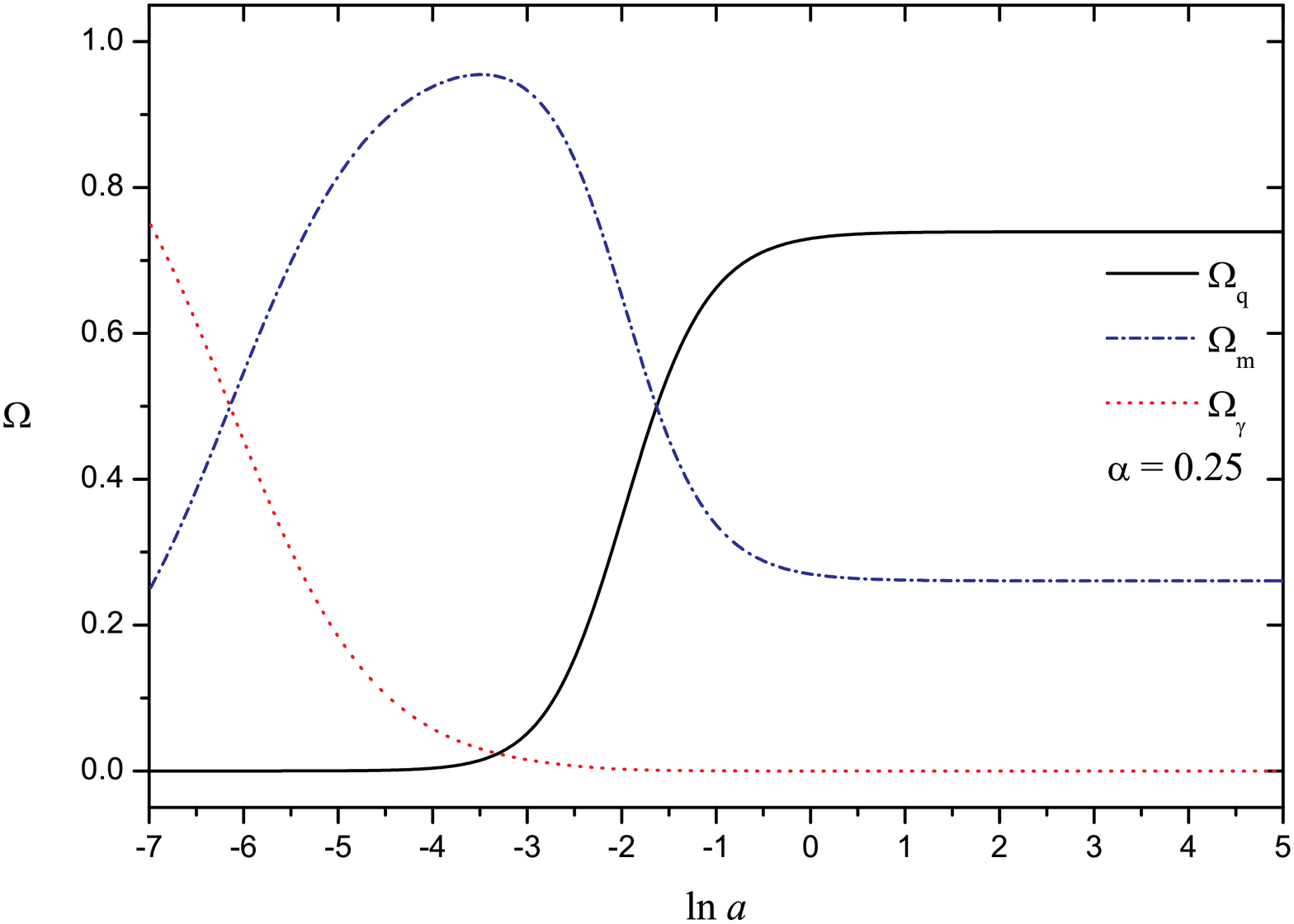}
\caption{Behavior of $\Omega_{\gamma}$ (dot line), 
$\Omega_{m}$ (dash dot line) and $\Omega_{q}$ (solid line) as a function of  $N=\ln a$ for $n=3,\;\beta = 0.1$ and $\alpha$ as indicated.}
\label{fig:2}
\end{figure}

In figure \ref{fig:2}, we show the evolution of $\Omega_q, \Omega_{m}$ and $\Omega_{\gamma}$ for
 different model parameters  $\alpha$ in the case of
 $w=0$ and $\beta = 0.1.$ It is easy to see that for the fixed  $\beta,$ which describes the decay dark matter  to  agegraphic dark energy,  increasing the value of the parameter  $\alpha$ leads to increase in the contribution of the relative density of dark matter $\Omega_{m}$.

\begin{figure}
\centering
\includegraphics[width=0.5\textwidth]{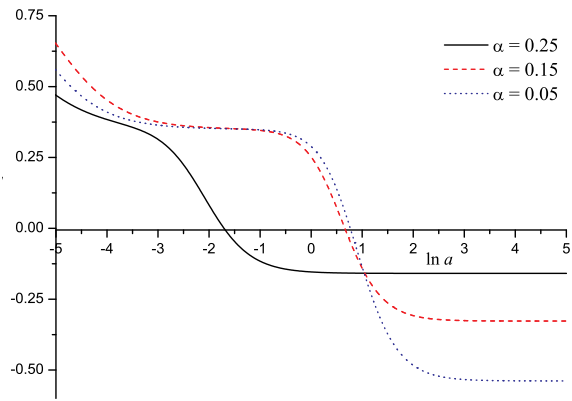}
\caption{Behavior of deceleration parameter  $q$  as a function of $\ln a$ for $\beta = 0.1$ and $\alpha$ as indicated.}
\label{fig:3}
\end{figure}
In figure \ref{fig:3}  we show the evolution of the deceleration parameter $q$  as a function of  $N=\ln a$  for $\beta = 0.1$ and $\alpha$ as indicated. 
The key distinction of this model is that the accelerated expansion of the Universe at late times is inherent even in the case when the phase of dark energy dominance does not occur.

\section{The model of interacting dark energy with a transient acceleration phase}\label{ISFA_TA}
Recent investigations seem to favor the cosmological dynamics according to which the accelerated expansion of the Universe may have already peaked and is now slowing down again \cite{Starobinsky}. As a consequence, the cosmic acceleration may be a transient phenomenon. 
One of the first on the possibility of a transient acceleration drew attention  J. D. Barrow \cite{Barrow}.  In his work he showed that in many well-motivated scenarios the observed spell of vacuum domination is only a transient phenomenon. Soon after acceleration starts, the vacuum energy's anti-gravitational properties are reversed, and a matter-dominated
decelerating cosmic expansion resumes. This was achieved by considering the dark energy in form of scalar quintessence field $\varphi$ defined by its self-interaction potential $V(\varphi).$ The potential of the scalar field in this case is the product of exponential and quadratic potentials.

The possible cosmological roles of exponential potentials have been investigated before, but almost always as a means of driving a period of cosmological inflation~\cite{LM85,pplane}. This requires potentials that are much flatter than those usually found in particle physics models.

In this section, we follow the latter route and investigate a new quintessence scenario driven by a rolling homogeneous scalar field with exponential potential $V(\varphi)$ interacting with dark matter on the agegraphic background. This  scenario   predicts  transient accelerating phase.

To describe the dynamic properties of such a Universe will adapt the system of equations (\ref{sys_xyz}) for this model. We introduce the modified variables:
\begin{eqnarray}  
x=\frac{\dot{\varphi}}{\sqrt{6}M_pH},\quad
y=\frac{1}{M_pH}\sqrt{\frac{V(\varphi)}{3}},\quad z=\frac{1}{M_pH}\sqrt{\frac{\rho_{m}}{3}},\quad  u=\frac{1}{M_pH}\sqrt{\frac{\rho_{_{q}}}{3}}.\label{var2}
\end{eqnarray}

The evolution of the scalar field is described by the Klein--Gordon equation, which in the presence of matter couplings is given by
\begin{equation}
\ddot{\varphi}+3 H \dot{\varphi} + \frac{dV}{d\varphi}=-  \frac{Q}{\dot{\varphi}}.
\label{eq:kgeqn}
\end{equation}
In this  section, we consider interactions $Q$ that are linear combinations of the scalar field and pressureless matter:
\begin{equation}
  \label{Q1}
  Q= 3H(\alpha\rho_\varphi+\beta \rho_m),
\end{equation}
where $\alpha$, $\beta$  are  constant parameters.

Without specifying the potential of the scalar field $V(\varphi),$ we obtain the system of equations in the form

\begin{eqnarray} 
x'&=& \frac{3x}{2}g(x,z,u)-3x + \sqrt{\frac{3}{2}}\lambda y^2-\gamma,\nonumber\hfill\\
 \label{sys_xyz}
y'&=&  \frac{3y}{2}g(x,z,u)- \sqrt{\frac{3}{2}}\lambda xy, \hfill\\
z'&=& \frac{3z}{2}g(x,z,u)-\frac{3}{2}z  + \gamma\frac{x}{z},\nonumber\hfill\\
u'&=& \frac{3u}{2}g(x,z,u)-\frac{u^2}{n},\nonumber\hfill
\end{eqnarray}

where 
$
g(x,z,u) = 2x^2+ z^2+ \frac{2}{3n}u^3
$
and
\begin{equation}
\gamma\equiv  -\frac{Q}{\sqrt{6}M_pH^2\dot{\varphi}},~\lambda\equiv  -\frac{1}{V}\frac{dV}{d\varphi} M_p.
\end{equation}
In these variables, we obtain

\begin{eqnarray}
  \nonumber
    Q&=&9H^3M_p^2\left[\alpha(x^2+y^2)+\beta z^3\right],\\ 
\gamma& = &\frac{\alpha(x^2+y^2)+\beta z^3}{x} \label{Q_gamma_xyz}. 
\end{eqnarray}

We will consider a scalar field with an exponential potential energy
density 
\begin{equation}\label{V_exp}
    V=V_0\exp\left(\sqrt{\frac{2}{3}}\frac{\mu\varphi}{M_p}\right),
\end{equation}
where $\mu$ is a constant.
In this case we obtain
\begin{eqnarray} 
x'&=& \frac{3x}{2}\left[g(x,z,u)- \frac{\alpha(x^2+y^2)+\beta z^2}{x^2}\right]-3x - \mu y^2,\nonumber\hfill\\
 \label{sys_xyz_V}
y'&=&  \frac{3y}{2}g(x,z,u)+\mu xy, \hfill\\
z'&=& \frac{3z}{2}\left[g(x,z,u)+\frac{\alpha(x^2+y^2)+\beta z^2}{z^2}\right]-\frac{3}{2}z ,\nonumber\hfill\\
u'&=& \frac{3u}{2}g(x,z,u)-\frac{u^2}{n}.\nonumber\hfill
\end{eqnarray}
In this model the deceleration parameter has the form
\begin{equation}
q=-1+\frac 32\left[2x^2+z^2+\frac{2}{3n}u^3\right]. 
\end{equation}
Note that in this model, the cosmological parameters  are not  explicitly  depend of the parameters of interaction, but only determine by the behaviors of dynamical variables. This fact complicates the analysis of our model. 

The fact that $q$ is independent of the interaction term implies  that the region of phase space is the same for all of the models considered. 
It is possible to make some qualitative comments about the system
(\ref{sys_xyz_V}) for some special cases.

Initially, considering the case, $y = 0,$  we  obtain the constraining relation, which should be imposed on the parameters of interaction for the emergence of critical points. It is easy to show that in this case, the condition of real energy density requires
\begin{equation}
\label{cond}
    2\sqrt{\frac{\beta}{\alpha}}>1+\alpha+\beta,
\end{equation}
which necessarily requires $0>\beta>\alpha,|\alpha|+|\beta|<1.$ 
This critical point corresponds to the matter-dominated Universe and is unstable. The cases of several such points are possible, but they are of no interest.
Finally, it can be shown that any of this equilibrium point within (but not on) the boundary will exist for $x_0<0.$  For $z\neq 0$ and constrain satisfied (\ref{cond}), it
\begin{equation}
    x_c=\left[\left(a+\sqrt{{\beta}/{\alpha}}\right)^{1/2}+a\right]z_c,
\end{equation}
 where $a=\left(2\sqrt{{\beta}/{\alpha}}-(1+\alpha+\beta)\right)^{1/4}.$

\subsection{Review of the case $Q=0$}
In this subsection we  consider in more detail the case of  absence of interaction between the scalar field and dark matter.
Critical points of the system (\ref{sys_xyz_V}) in this $(\alpha=\beta=0)$ case are given in the Table \ref{tab:2}.
As can be seen from the table, the system (\ref{sys_xyz_V}), there are six physically admissible critical points, the latter of which is the attractor.
The first critical point, $(1,0,0,0),$ is unstable and  corresponds to the scalar field  dominated era with extremely rigid equation of state, the second critical point corresponds to the period of evolution when the scalar field behaves as a cosmological constant.
The other point, $(0,0,1,0),$ is physically unrealistic. It corresponds to a Universe filled with  dark matter (it contains neither scalar field, nor agegraphic dark energy) and it is also unstable.
Fourth critical point $(0,0,0,1)$  correspond to the Universe consisting  only of the  agegraphic dark energy and has already been discussed in detail. 

The physical interest present the final sixth critical point, which is the an attractor. It corresponds to the Universe consisting of a scalar field and agegraphic dark energy.
The location of this critical point is completely determined by the parameter of the potential $\mu$ and the value of $n:$
\begin{equation}
    \begin{array}{cccccl}
   x_{*}&=&\frac{2}{3n\mu}u_{*}, \quad& y_{*}&=&\sqrt{1-\left(1+\frac{4}{9n^2\mu^2}\right)u_{*}^2},\\  
	z_{*}&=&0, \quad						& u_{*}&=&\frac{3}{2n\mu^2}\left(-1+\sqrt{1+\frac{4n^2\mu^4}{9}}\right).
\end{array}
\end{equation}

The fact that $x_*\propto u_*$ is the characteristic feature of the tracking solutions. Note also that between the scalar field and agegraphic dark energy, evidence of background interactions. There is because the dynamics of the scalar field affects agegraphic dark energy. This effect is that agegraphic dark energy, having a negative pressure, affect on the rate of expansion of the Universe, which affects the Hubble parameter, which is included in the Klein-Gordon equation for a scalar field.

\begin{table}
\caption{Location of the critical points of the autonomous system of Eqs. (\ref{sys_xyz_V}), their stability and dynamical behavior of the Universe at those points.}
\label{tab:2}
\begin{tabular}{ccccc}
\hline\noalign{\smallskip}
$(x_c,y_c,z_c,u_c)$ & Stability & $q$ & $w_{\varphi}$& $w_{tot}$\\
 coordinates &  character  & & & \\
\noalign{\smallskip}\hline\noalign{\smallskip}
$\;(1,0,0,0)$ &  unstable  & $2$ &$ 1$& $1$\\
$\;(0,1,0,0)$   & unstable  & $-1$ &$-1$& $-1$\\
 $\;(0,0,1,0)$ &unstable &$\frac{1}{2}$\quad &$\nexists$&0\\
 $\;(0,0,0,1)$ & stable  & $-1+\frac{1}{n}$ &$ \nexists$& $-1+\frac{2}{3n}$\\
 $(-\frac{3}{2\mu},\frac{3}{2\mu},\sqrt{1-\frac{3}{2\mu^2}},0)$ & unstable &\quad$\frac{1}{2}$ &$\nexists$&0\\
 $(x_*,y_*,0,u_*)$ & attractor &$q_*<0$&$w_{\varphi*}$&$w_{tot*}$\\
\noalign{\smallskip}\hline
\end{tabular}
\end{table}

The attractor, once reached, brings to zero the matter density. To allow for the observed matter content of the Universe, we have to select the initial conditions, if they exist, in such a way that the attractor is not yet reached at the present time, but the expansion is already accelerated.

For example, consider the case when $\mu = -3, n=3.$ The dynamics of such a Universe is consistent with the observed data, which will be shown in the following section.  We note only that in this model the values of deceleration parameter and the scalar field state parameter in the attractor are equal respectively $q_*\approx -0.68,~w_{\varphi*}\approx -0.78.$

\subsection{Review of the case $Q=3H\alpha\rho_\varphi$}
In the above case, the phenomenon of transient acceleration that occurs in such a Universe does not match the observations.
However, it is possible to fit the observational data with this model in the case when dark matter and scalar field interact. In this subsection we consider the special case of interaction (\ref{Q1}) when $ \beta = 0.$

\begin{figure}[t]
\centering
\includegraphics[width=0.45\textwidth]{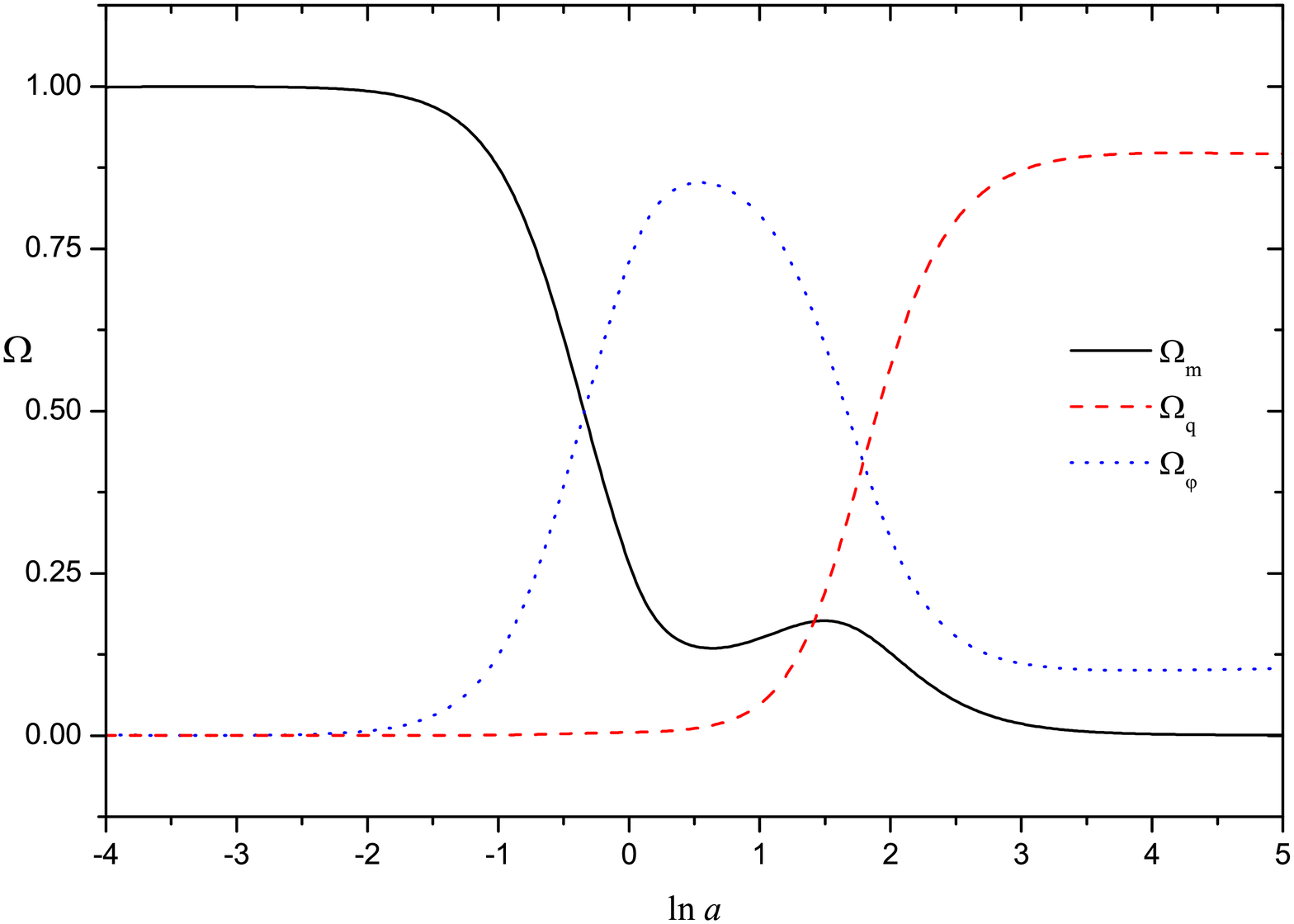}
\includegraphics[width=0.45\textwidth]{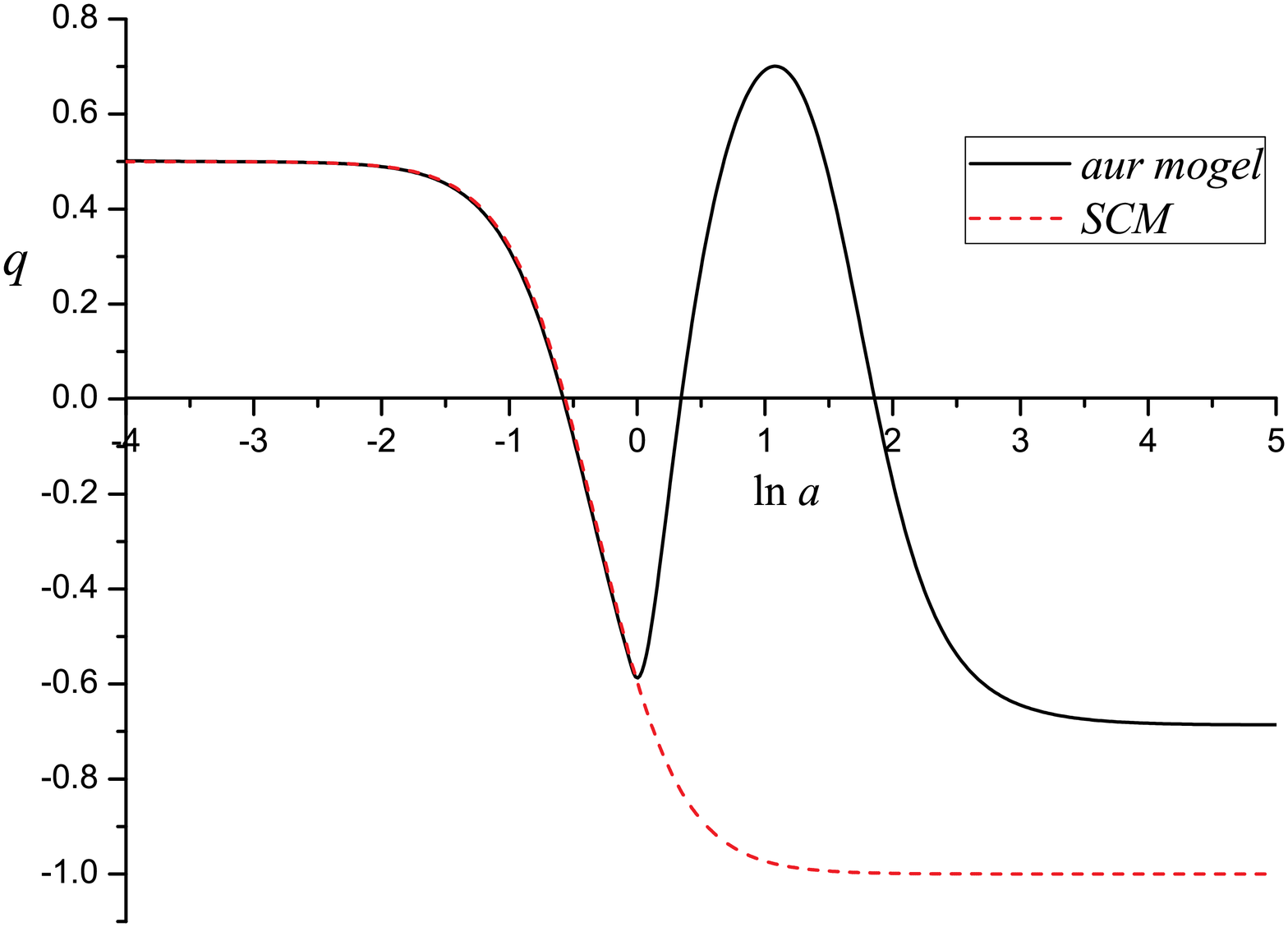}
\caption{Behavior of $\Omega_{\varphi}$ (dot line), 
$\Omega_{q}$ (dash  line) and $\Omega_{m}$ (solid line) as a function of  $N=\ln a$ for $n=3,\;\alpha = 0.005$ and $\mu = -3$ (left side). Evolution of deceleration parameter for this model (right side).}
\label{fig:4}
\end{figure}

In figure \ref{fig:4}, we show the evolution of $\Omega_q, \Omega_{m}$ and $\Omega_{\varphi}$ for  cosmological model  in the case when $\alpha=0.005$,
 $\mu=-3$ and $n = 3.$

From the form of equations and the character of the interaction it can be easily understood that neither the nature nor the location of the attractor, that have been found in the previous subsection dos not change with the inclusion of the interaction. Interaction only affects the behavior of dynamical variables  that are the correspond to trajectory  in phase space  between critical points. This is a consequence of the above degeneracy from the parameters of interaction.

With these values of the parameters of interaction, transient acceleration begins almost in the present era. In our model as in most cosmological models, where the scalar field plays the role of dark energy it begins to dominate causing a period of accelerated expansion of the Universe. Its consequence of the accelerated expansion of the Universe contribution of ADE increases, resulting that the background (space) is changing faster than the field and it becomes asymptotically free. This field has extremely rigid equation of state that leads to the fact that the accelerated expansion of the Universe is slowing down. Soon, however, when the contribution of ADE has grows enough so that the scalar field cannot longer impede the expansion of the Universe, it begins to accelerate again.

\section{Observational data}\label{OBS}
In the present section, we will consider the latest  cosmological observations, namely, the 397 Constitution SNIa  dataset~\cite{data_1}.

The data points of the 397 Constitution SNIa compiled  in~\cite{data_1} are given in terms of the distance modulus  $\mu_{obs}(z_i)$. On the other hand, the theoretical  distance modulus is defined as
\begin{equation}
     \mu_{th}(z_i)\equiv 5\log_{10}D_L(z_i)+\mu_0\,,
\end{equation}
where $\mu_0\equiv 42.38-5\log_{10}h$ and $h$ is the Hubble  constant $H_0$ in units of $100~{\rm km\,s^{-1}\,Mpc^{-1}}$, whereas
\begin{equation}
     D_L(z)=(1+z)\int_0^z \frac{d\tilde{z}}{E(\tilde{z};{\bf p})}\,,
\end{equation}
in which $E\equiv H/H_0$, and ${\bf p}$ denotes the model parameters. 

\begin{figure}[t]
\centering
\includegraphics[width=0.45\textwidth]{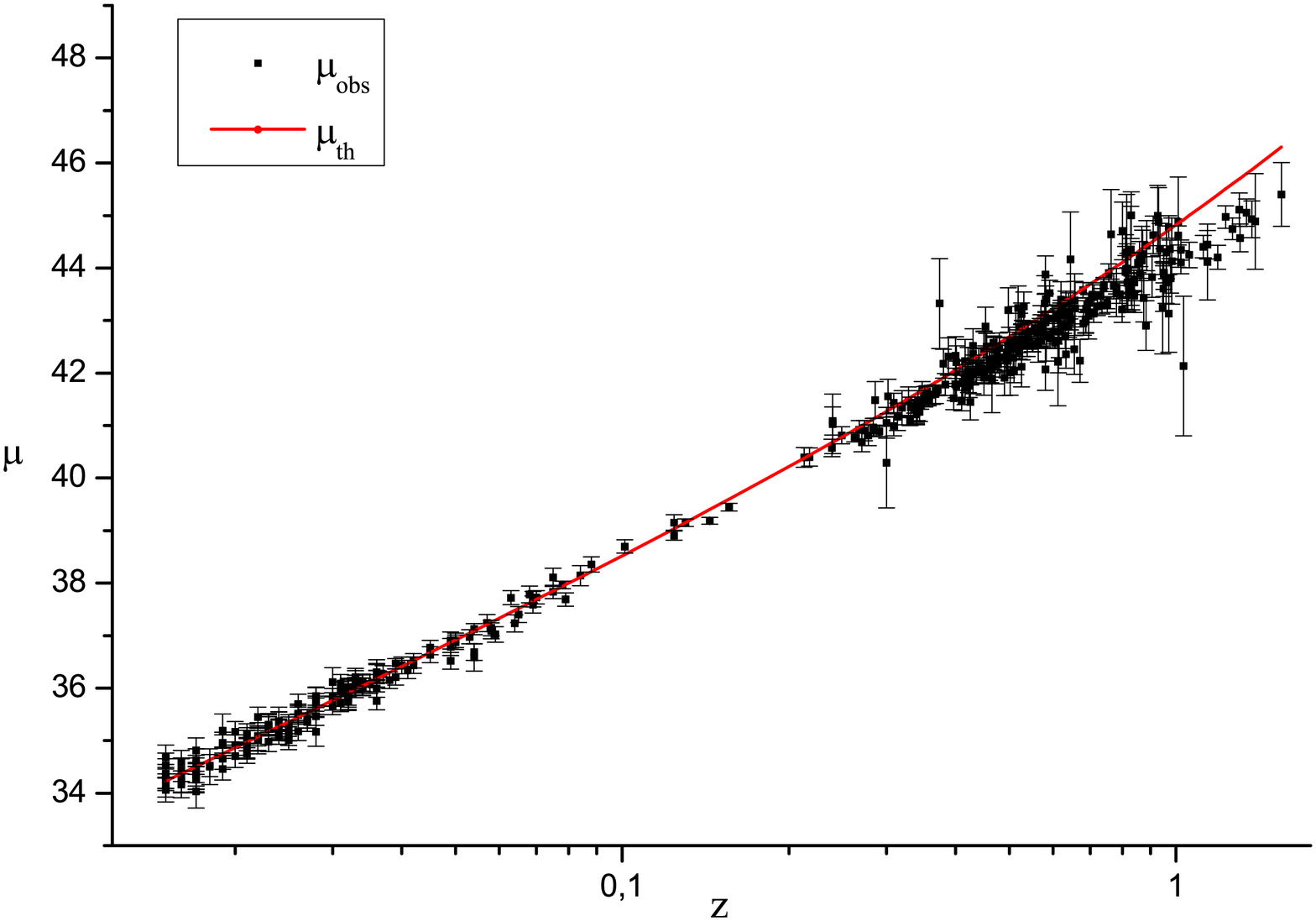}
\includegraphics[width=0.45\textwidth]{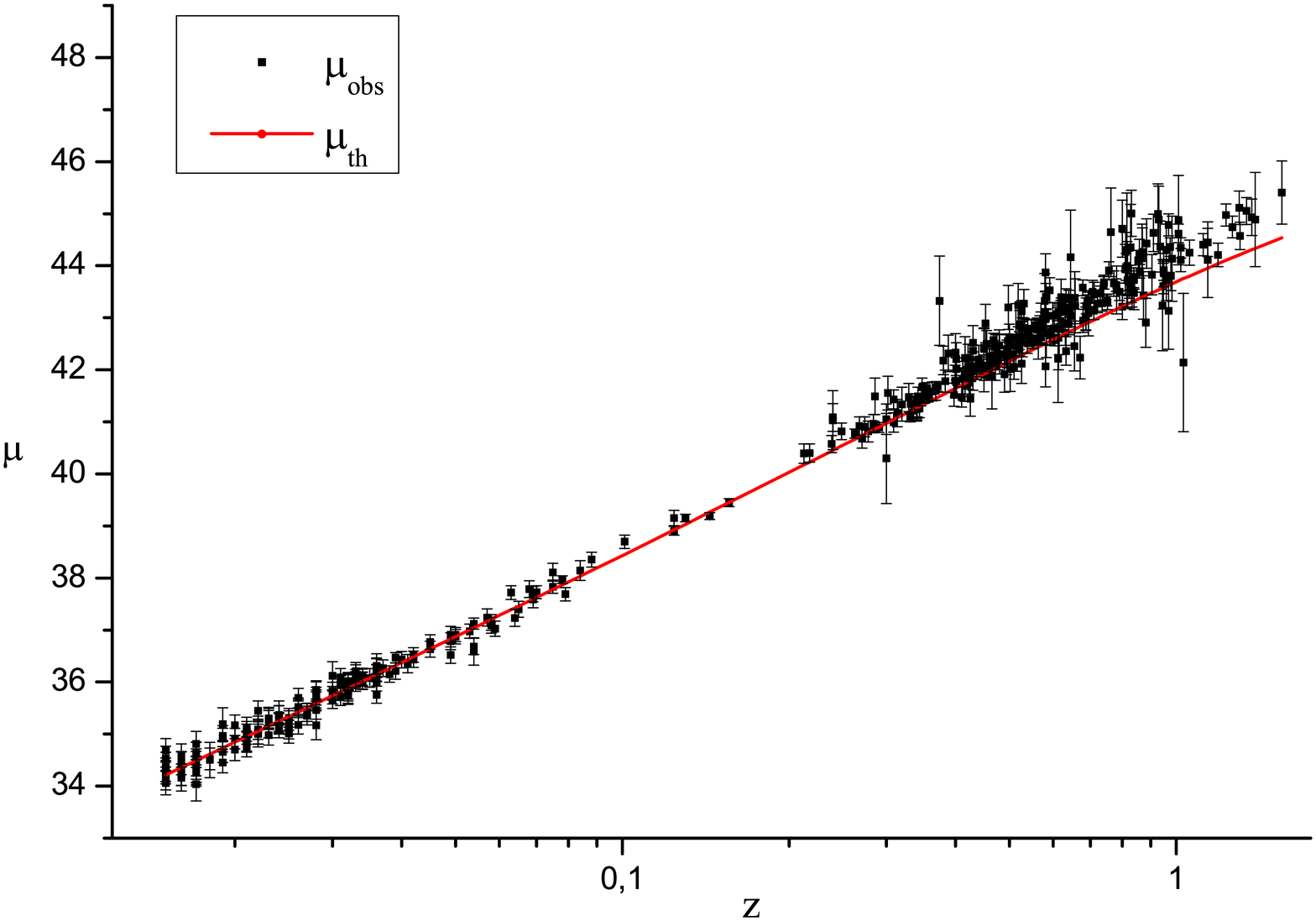}
\caption{The dependence of the modulus distance from the redshift, theoretically calculated (solid red line) by the model with $Q=3H(\alpha \rho_q +\beta\rho_m$), $\alpha=0.25$, $n=3$ (left side)  and $Q=3H\alpha \rho_\varphi$, $\alpha=0.005$, $\mu=-3$ and $n=3$ (right side).  $ h = 0.65 $ $\rm \, km  \,c^{-1}\,Mpc^{-1} $ for both models  and found the results of observations of supernovae type 1a \cite{data_1}}.
\label{fig:mu_z_1}
\end{figure}

Theoretical distance modulus  will be different for the different models and comparing $\mu_{th}(z_i)$ with $\mu_{obs} (z_i) $, one can judge the plausibility of an cosmological model. So  to understand whether the theoretical model corresponds to the  observational data is enough to know the value of $E\equiv H/H_0$, which is easy to find through a system of equations (\ref{sys_xyz_V}). As seen from figure \ref{fig:mu_z_1} our  models are in accordance with the observational data. 

\section{Conclusion}
The original agegraphic dark energy model was proposed in~\cite{0707.4049} based  on the K\'{a}rolyh\'{a}zy uncertainty relation, which arises  from quantum mechanics together with general relativity.  The interacting agegraphic dark energy model has certain advantages compared to the original agegraphic or holographic dark energy model. Many studies show that this model gives an opportunity to explain the accelerated expansion of the Universe without a cosmological constant or some form of the scalar field. All the three models give dynamics of the Universe which are virtually indistinguishable from SCM, but without most of its problems, such as the cosmological constant, fine tuning and coincidence problems.

Some authors have recently suggested that the cosmic acceleration have already peaked and that we are currently observing its slowing down \cite{Barrow,Starobinsky,Lima}.
Under a kinematic analysis of the most recent SNe Ia compilations, the paper \cite{Lima}  shows  the existence of a considerable probability in the relevant parameter space that the Universe is already in a decelerating expansion regime. 

One of the deficiencies of original ADE model is the inability to explain the phenomenon of transient acceleration. 

Density of holographic dark energy is determined by the surface terms in action, while volume terms are usually ignored. We take into account both surface and volume terms, where the latter correspond to (described by) homogeneous scalar field with exponential potential $V(\varphi)$.

We consider a model of Universe consisting of dark matter interacting with a scalar field on the agegraphic background. It is shown that this model can explain the transient acceleration. This model also is in accordance with the observational data.

\section*{Acknowledgements}

We are grateful to our research supervisor  Prof. Yu.L. Bolotin for kind help and discussions. We also thank V.A. Cherkaskiy for careful reading and editing of this article.

\bibliographystyle{JHEP}

\end{document}